# A local view of the laser induced magnetic domain dynamics in CoPd stripe domains at the picosecond time scale.

V. López-Flores[1], M.-A. Mawass[1,2], J. Herrero-Albillos[2,3], A.A. Uenal[2§], S. Valencia[2], F. Kronast[2], and C. Boeglin[1]*

[1] Université de Strasbourg, CNRS, Institut de Physique et Chimie des Matériaux de Strasbourg, UMR 7504, F-67000 Strasbourg, France.

[2] Institut für Methoden und Instrumentierung der Forschung mit Synchrotronstrahlung Helmholtz-Zentrum Berlin für Materialien und Energie GmbH, Albert-Einstein-Str. 15, 12489 Berlin, Germany

[3] Centro Universitario de la Defensa, Zaragoza, Spain.

\* *Corresponding author:*

*mail:* *christine.boeglin_@ipcms.unistra.fr*

*Address:* Institut de Physique et de Chimie des Matériaux de Strasbourg (IPCMS)

23, rue du Loess 67034 Strasbourg

§ *Now at Max-Born-Institut, Berlin, Germany.*

**Abstract**

The dynamic of the magnetic structure in a well ordered ferromagnetic CoPd stripe domain pattern has been investigated upon excitation by femtosecond infrared laser pulses. Time-resolved X-ray magnetic circular dichroism in photoemission electron microscopy (TR-XMCD-PEEM) is used to perform real space magnetic imaging with 100 ps time resolution in order to show local transformations of the domains structures. Using the time resolution of the synchrotron radiation facility of the Helmholtz-Zentrum Berlin, we are able to image the transient magnetic domains in a repetitive pump and probe experiment. In this work, we study the reversible and irreversible transformations of the excited magnetic stripe domains as function of the laser fluence. Our results can be explained by thermal contributions, reducing the XMCD amplitude in each stripe domain below a threshold fluence of 12 mJ/cm$^2$. Above this threshold fluence, irreversible transformations of the magnetic domains are observed. Static XMCD-PEEM images reveal the new partially ordered stripe domain structures characterized by a new local magnetic domain distribution showing an organized pattern at the micrometer scale. This new arrangement is attributed to the recovery of the magnetic anisotropy during heat dissipation under an Oersted field.

# 1. Introduction

Femtosecond laser excitation on ferromagnetic films leads to a macroscopic ultrafast demagnetization on a sub-picosecond time scale [1]. This pioneering work opened the field of femtomagnetism extending nowadays to several aspects as discussed in literature [1, 2, 3, 4, 5, 6, 7, 8]. Among them the dynamic response of magnetic domains to such excitations and the laser induced switching of the magnetization [5, 9] stands out. During the past years the study of the dynamics of the spins in magnetic domains have motivated many studies addressing different aspects of the dynamical processes, discussing Elliot Yafet or thermal effects as well as Circular dichroism effects as possible mechanisms to understand the switching of spins. However, complex magnetic domain transformations triggered by laser excitations have been studied by several groups [5, 9, 10 - 13] and show that the ultrafast switching dynamics are far from being understood. Most of these experiments have been performed by magneto-optic Kerr effect. Even though time resolutions down to 50 fs can be achieved by means of ultrafast Kerr microscopy, this technique shows a limited lateral resolution [14]. This drawback can be overcome by using synchrotron-based techniques within the soft X-ray energy range combining magnetic and element-selectivity with ~ 10 nm spatial resolution. Recent advances on time-resolved X-ray techniques (synchrotrons, X-FEL and High Harmonic Generation (HHG) laser sources) have improved the understanding of the spin dynamics in single and multidomain structures [3, 6, 10, 11]. Techniques based on X-ray magnetic circular dichoism (XMCD) have recently provided new insights into the dynamics of spins upon laser excitations. [3, 6, 13]. However, combining the ultimate time (~ few picoseconds) and space resolution (~ few nanometers) in order to describe ultrafast magnetic domain dynamics is still a challenge. Several X-ray based techniques have been used to reveal different aspects of the micromagnetic domains dynamics with various time resolutions [10 - 13, 15 - 20]. We know that sub-picosecond time resolution using resonant magnetic X-ray scattering only provide spatially averaged information and therefore many local aspects of the domain organisation are still unknown. Moreover, when using X-ray holography techniques [19], the field of view is usually too small to capture specific and local dynamics in the magnetic domains. A recent study using time-resolved resonant magnetic X-ray scattering showed that IR pump pulses induce demagnetization and reorientation of the magnetic domains in CoPd stripe patterns [12]. This study could provide averaged parameters at the nanosecond timescale describing the recovery of the magnetic pattern after a single intense Infra-Red (IR) pump.

In this work we show that a real space view of the picosecond demagnetization dynamics in CoPd stripe domains can reveal new aspects of the local transformations of domains upon femtosecond laser excitations. Such study can be performed using X-ray photoemission electron microscopy (XPEEM) operated in a pump-probe configuration with few tens of picoseconds time resolution. It combines adequate spatial resolution of 30 nm in order to describe the local picosecond dynamics of the magnetic domains after laser excitation [16, 17]. Both the reversible and the irreversible transformations induced by the IR laser have been addressed in our study. We observe two different regimes of transformations of the domain structures below and above a threshold pump fluence of 12 mJ/cm$^2$ corresponding to the reversible pump conditions used for CoPd films in our previous studies [6, 7, 10, 12, 19].
Our results show, by exploiting the quantitative XMCD-PEEM images that the laser induced quenching of ferromagnetic stripe domains is characterized by a reduction of the magnetization inside the magnetic domains and probably by a broadening of the domain walls. We show that this transformation also involves a thermal induced partial transfer of the magnetization from out-of-plane to in-plane. The laser induced evolution of the magnetic stripe pattern is reproduced by micromagnetic simulations based on the temperature dependent anisotropy in CoPd. Moreover, we describe the new local magnetic domain configuration which is induced above 12 mJ/cm$^2$ pump laser fluence.

## 2. Experiment

### 2.1 Samples

CoPd magnetic films are deposited at 423 K on top of a conductive doped Si substrate by means of electron beam co-evaporation of Co and Pd from highly pure target metals. The 50 nm thickness of the layer and the alloy composition of $Co_{0.60}Pd_{0.40}$, ensured regular magnetic stripe domains and a large out-of-plane magnetic anisotropy in the film. Parallel stripes, showing typical lateral sizes of 80 nm, were induced in the samples by an in-plane demagnetization procedure. The maximum applied field was 1.6 T, and the reduction rate was 1% of the previous step, until 5 mT, where the field was reduced to 0. Due to the magnetic anisotropy, alternating out-of-plane stripe domains are formed with the stripes orienting parallel to the demagnetizing field [21]. Following this procedure many static and dynamic Magnetic Force Microscopy (MFM) or XMCD-based experiments have been conducted in the past [10, 12, 13, 20, 21- 24].

The base pressure during co-evaporation was $1 \times 10^{-9}$ mbar. A 20 nm Pd buffer layer was deposited to provide an efficient transfer of the pump induced heat. A 2 nm Pd + 1 nm Al capping layer was used to prevent oxidation of the magnetic layer. No oxidation could be evidenced at the Co $L_3$ edge during our experiment.

SQUID measurements were carried out to characterize the out-of-plane anisotropy of the samples. Both in-plane and out-of-plane magnetizations were measured as a function of the temperature to determine the temperature dependence of the magnetic anisotropy. As shown in Figure 1, the shape of the magnetization curves with the out-of-plane external fields is characteristic of the presence of magnetic domains. At 50 K these magnetization curves clearly show that the sample is more easily magnetized when the field is applied perpendicularly to the layer, indicating a preferred orientation of the magnetization out-of-the plane at low temperatures. The temperature dependent evolution of the in-plane and out-of-plane magnetization curves evidences the change of the magnetic anisotropy from out-of-plane to in-plane at higher temperatures. This temperature dependence is compatible with the temperature dependent anisotropy constant as given in literature for CoPd films [12, 22, 25, 26].

### 2.2 Experimental Configuration

X-ray photoemission electron microscopy (X-PEEM) measurements were carried out at the UE49/2-PGM-SPEEM beamline at BESSY II synchrotron (Helmholtz Zentrum Berlin). The experimental setup is described elsewhere [27]. The magnetic contrast on the X-PEEM is achieved by taking advantage of the sensitivity to ferromagnetic domain direction of the X-ray magnetic circular dichroism (XMCD) effect at the $L_{2,3}$ edges of transition metals. Accordingly, we collected X-PEEM images at the Co $L_3$ edge (778 eV) with both left and right handed circularly polarized light. Their subtraction divided by their sum leads to the XMCD asymmetry image, which contrast arises from magnetic domains only, and which intensity and sign depends on the projection of the magnetization along the incident X-ray direction.

The pump laser is a Ti:Sapphire laser system operating at 800 nm wavelength and 100 fs pulse width with a repetition rate of 5 MHz. The laser and the X rays impinge on the surface of the film with an incidence angle of 16º from the surface plane. By operating the microscope at a field of view of 5 μm, we ensure to image the surface of the CoPd film homogeneously excited by the IR laser. This is due to the fact that the laser spot on the sample is 60x20 μm$^2$ which is larger than the trace of the X-ray spot (30x10 μm$^2$). Moreover, the penetration depth of the IR laser for the CoPd alloy (10 nm) is larger than the XMCD-PEEM probing depth (2-5 nm), guarantying that the probed region is homogeneously excited in depth as well.

For the time-resolved pump-probe experiments we used the single-bunch filling mode of BESSY synchrotron, which produces a single X-ray pulse with a frequency of 1.25 MHz and width of about 60 ps. The experiment leads to an XMCD-PEEM image with an estimated time resolution of 60 ps. Additionally, the multi-bunch X-ray beam was used to characterize the static stripe pattern configuration before any laser excitation and after the series of pump-probe experiments.

During the time-resolved experiment, the laser pump pulse was synchronized with the X-ray probe pulse by a phase locked loop (PLL) circuit. The time overlap between X rays and IR laser pulse was detected exploiting the burst of secondary electrons at a defect resonating with the laser. An electromechanical trombone delay line was used to shift the phase of the storage ring master clock signal used for synchronization, defining the phase shift of the IR-laser and thereby the pump-probe delay.

Time-resolved XMCD-PEEM images were taken at room temperature at two different pump-probe delays: -250 ps and +190 ps. At these two delays, time resolved and time integrated XMCD-PEEM images were taken at different IR laser fluences. The sequences of images characterises the laser induced demagnetization and re-organisation of the magnetic stripe pattern.

Irreversible transformations of the stripe pattern have been observed by recording time integrated XMCD-PEEM images at different incident laser fluences between 8-18 mJ/cm2 until the irreversible damage threshold was reached (above 12 mJ/cm$^2$). We repeat this study for different in-plane directions of the magnetic stripe domains (35º and 90 °) in respect to the X-ray propagation direction.

## 3. Experimental results

*Dynamics of magnetic stripe domains.*

The XMCD-PEEM characterization previous to any laser exposure confirms the presence of magnetic stripes at room temperature shown in figure 2(a). The blue and red contrast results from the projection of the out-of-plane magnetization to the X-ray propagation direction. The FOV of the XMCD-PEEM image is 5 μm. As shown by the 2D Fourier Transform (FT) of the PEEM image of magnetic domains the presence of two spots evidences the periodicity of the ordered magnetic stripes perpendicular to the X-ray incidence.

Time-resolved XMCD-PEEM images recorded at -250 ps before (left) and at +190 ps after the laser excitation is shown in Figures 2b-g for different laser fluences F (F= 8.8; 9.5 and 10.7 mJ/cm2). The left column shows the transient PEEM images before arrival of the laser pump at a delay of -250 ps, whereas the right column shows the transient images after the laser excitation at a delay of +190 ps. The FOV of the XMCD-PEEM images is 5 μm. In Figure 3 we show the corresponding FT images of the different XMCD-PEEM images recorded at the pump-probe delays of -250 ps (left column) and +190 ps (right column). The green circles show the limits used for the azimuth integration defining the intensity distributions I(Φ).

Due to the repetitive pump-probe, an IR fluence dependent DC heating increases the mean XMCD values at any delays. This corresponds to the increase of the in-plane component of the magnetization as measured by the grazing incident X rays and is confirmed by the temperature dependent SQUID data. Note that under our experimental pump-probe conditions we estimate that the increase of the sample's temperature due to the laser is of about 50 K at -250 ps and ~100 K at +190 ps. In order to visualize the typical effects of the DC heating on our time-resolved XMCD-PEEM images, we plot in Figure 4 the 3 qualitative line profiles representing the local XMCD values and the temperature induced offsets across the stripe domains. The colour code (blue-white-red) qualitatively reproduces the colour codes observed in our XMCD-PEEM images and is proportional to the normalized XMCD. The drawn continuous lines (black, green, and orange) evidence the local magnetization across 3 successive stripe domains and domain walls. Our sketch qualitatively describes typical line profiles perpendicular to the stripe direction and shows 3 different stripe domain configurations: Without laser (black line), with laser at a delay of -250 ps (green line) and with laser at a delay of +190 ps (orange line). A XMCD offset shift towards higher XMCD values is shown by green and orange dashed lines for both delays -250 ps and +190ps. The offset value is given by the average XMCD through the domain pattern and is proportional to the DC heating.

The orange curve reproducing the magnetization across the domain wall at positive delay of +190 ps reproduces the well known enlarged domain wall. This line profile follows qualitatively the Object Oriented MicroMagnetic Framework (OOMMF) simulations which are shown in Figure 5.

The simulated magnetic line profiles have been obtained for a 50 nm thick CoPd alloy film for two different and realistic values of the magnetic anisotropy constants defining the system at room temperature and at an elevated temperature of ~400 K, $K_m$=5e6 J/m$^3$ and $K_m$=5e5 J/m$^3$, respectively. The magnetic parameters for the CoPd alloy have been taken from literature [12, 22, 25, 26]. The OOMMF simulated line profiles in Figure 5 evidence the change of the domain wall between two up and down domains. Comparing the two simulated profiles we identify the effect of temperature on the domain walls. As expected, we obtain wider domain walls at elevated temperature. In our simulations, the domain walls show sizes of ~3nm to ~15 nm when the anisotropy is divided by 10. Unfortunately, this modification cannot be imaged with XMCD-PEEM, since it is below the experimental spatial resolution of 30 nm.

Furthermore, at a given pump fluence, the XMCD offset (Figure 4) is slightly stronger for the delays at +190 ps than for -250 ps (orange compared to green dashed lines). This is present at any IR fluences and corresponds to the time-dependent increase of the in-plane component of the magnetization. Table 1 shows the mean XMCD values extracted from the PEEM images, taking into account the geometry of the experiment. The extracted value of the reference image (Figure 2a) without laser is close to 0% corresponding to an equal number of up and down magnetic domains. At a fluence F=8.8 mJ/cm$^2$ and at delay -250 ps the DC heating induces an increase of the mean XMCD value by 10% which noticeably decreases for higher IR pump fluences (F= 9.5 and 10.7 mJ/cm$^2$). At moderate IR fluence of F=8.8 mJ/cm$^2$ we interpret the large XMCD mean value as related to a sizable heat induced in-plane component combined with a lower out-of-plane magnetization contrast (green line in sketch). At larger IR fluences of F = 9.5 and 10.7 mJ/cm$^2$ the mean XMCD decreases by -50 % and -65 % respectively. Such a significant drop can be interpreted by temperature induced demagnetization along both, the in-plane and out-of-plane direction.

As can be seen from Table 1, the mean XMCD values at a delay of +190 ps are consistently higher for all images compared with their counterpart at -250 ps. The +2% increase of the mean XMCD values at t = +190 ps reflects a time-dependent thermal induced magnetization rotation from out-of-plane to in-plane, driven by a lower magnetic anisotropy at +190 ps. This is indeed verified by comparing the local XMCD amplitudes of the magnetic stripe domains at -250 ps and +190ps as sketched in Figure 4 (orange line and orange dashed line).

To quantify the reduction of the XMCD amplitude (proportional to magnetisation in the up and down magnetic domains) we performed the FT of the XMCD-PEEM images as shown in Figure 3. Using the FT we are able to focus and analyze the changes in the spatial distribution of the observed intensities. The azimuthal intensity distribution I(Φ) shown in Figure 6, yields a characterization of the ordering and the correlation length of the magnetic domains. However, because of the poor FT data quality, related to the limited 5 μm spatial integration, we will limit our analysis to the normalized azimuthal scattering intensity distributions as shown in figure 6 (using a 2 degree step).

The I(Φ) have been extracted from the FT images delimited by the two green circles as shown in Figure 3. We normalize I(Φ) by the background level of the FT image. In this way we can exploit the peak intensities and their half maximum to characterise the average magnetization in the domains and their correlation length. Figure 6, shows how the XMCD amplitudes are slightly reduced for all FT -XMCD-PEEM images at +190 ps, corresponding to the transient laser induced reduction of the magnetization in the domains (Figure 6 at fluence F = 8.8, 9.5 and 10.7 mJ/cm$^2$)). At these delays, we evidence the transient demagnetization of -30% and -50 % in the up and down oriented magnetic strip domains, which are mainly due to thermal excitations. Indeed at delays of ~ 100 - 500 ps, it is well known that the system has not yet reached the recovery [6, 12]. At -250 ps delay (continuous lines) the reduction of the peak intensities with increasing fluences illustrates the effect due to DC heating on the magnetization of the stripe domains. However no discrimination between in-plane and out-of-plane spin directions is possible.

Furthermore, we observe in Figure 6 a change in the full width at half maximum (FWHM) from 55° without laser (black line), down to 40° for the other cases. The reduced FWHM evidences a larger correlation length of the magnetic domains. In the XMCD –PEEM images one could also notice less labyrinth-like domains after exposure to the laser (Figure 2b, c, d, f). No significant differences are observed between -250 ps and +190ps, confirming that this effect is linked to the DC heat induced reorganisation of the stripe pattern.

*Domain reorientation.*

In order to describe the irreversible transformations of the stripe domains at higher laser fluences we have studied the change of the static XMCD –PEEM images obtained without and with the pump laser on the surface of the film. The static imaging mode has been used for different fluences of 14.7 - 18.1 mJ/cm$^2$, ensuring that the previously defined threshold fluence of 12 mJ/cm$^2$ was exceeded. In order to describe in-plane components of the spin as well as the stripe domains themselves we recorded and compared two different sample azimuths in respect of the incoming X rays. In the first one we orient the stripes perpendicularly to the X-ray incidence (Figure 7a) and in the second, we orient the stripes nearly parallel (~35°) to the X rays (Figure 7c). The FOV of the images in Figure 7 is 5 μm. In this last sample orientation, both the out-of-plane and in-plane spin components (projections) are visible whereas in Fig. 7a only the up and down spins are visible. We therefore conclude that the magnetization of the sample is not purely out-of-plane, but is slightly tilted along the stripe direction at room temperature. We again performed the FT of the images in order to highlight the stripe orientations. In the FT image of Figure 7c (inset) we observe that due to the contributions of the in-plane projections, an additional structure with a non-symmetrical spiral shape appears. This structure is absent in the FT image in Figure 7a (inset).
After having exposed the CoPd film surface to the laser, we performed XMCD-PEEM imaging at room temperature for both orientations. The results are shown in Figure 7b and 7d. We note that the two magnetic domain images recorded without pump laser, at room temperature are obtained at different sample positions.
In Figure 7b we evidence that nearly half of the field of view on the right side of the image shows an almost uniform contrast corresponding to a magnetic domain that have irreversibly turned the magnetization to the in-plane direction with a strong component parallel to the X rays (red background). Over the complete field of view one can notice a new stripe ordering, confirmed by the FT shown in the inset of Figure 7b. The new alignment of the stripe domains is no more perpendicular to the X-rays incident direction and allows now to observe the in-plane components of the spins together with the stripe structures. Such a re-orientation process has already been shown and explained in our recent work [12]. However we have here access to the local view of the new domain organisation as well as in-plane projection of the spins. We can also confirm the re-orientation along a given angle of 60º in respect to the X-ray incidence induced by large pump fluences. By performing the same experiment with the second orientation of the stripes at 35° to the X rays we are able to reproduce this final state of the domain structure and reorientation (Figure 7d). We conclude that the final domain structure is independent from the initial orientation of the stripe domains. The reorientation can be confirmed in the FT diagram depicted in the corresponding insets where besides a 60º new alignment from the X-ray incidence, we also observe a weakening of their intensities. The orientation of the stripes can be analysed in more detail using the azimuthal intensity distribution I($\Phi$) as shown in Figure 8. In Figure 8a (black line), we follow I($\Phi$) for the initial orientation of the stripes perpendicular to the X rays (corresponding to Figure 7a) and his transformation after the laser excitation. In Figure 8b (black line) we plot I($\Phi$) for the second initial orientation of the stripes, at 35º to the X rays and his transformation after the laser excitation. We notice that the intensity of the FT spots (maximum of I($\Phi$)) for the original perpendicular orientation is more intense than the others due to the negligible contribution of the in-plane components. Moreover, the laser induced reorientation shows reduced I($\Phi$) intensities and a common final reorientation angle of 60º from the X rays (Figure 8 a, b).

Comparing both cases, we can conclude that the reorientation does depend neither on the sample original geometry nor on the incidence plane of the laser or X-ray. Thus, we believe that an external cause, likely a weak symmetry breaking external magnetic field, might be causing this preferential reorientation. During the recovery of magnetization at a given sample temperature, a weak magnetic anisotropy appears, allowing for a partial orientation of the domains with respect to the weak external magnetic field. Such a mechanism was confirmed by micromagnetic simulations in our previous work [12].

## 4. Conclusions

The dynamic of the magnetic stripe domain structure have been investigated upon excitation by femtosecond infrared laser pulses. Time-resolved XMCD-PEEM reveals the transient magnetic domains as well as final domain structures at room temperature. We studied the reversible and irreversible transformations of the excited magnetic stripe domains as a function of the laser fluence. We show that at moderate laser fluences below 12 mJ/cm2, thermal contributions reducing the XMCD amplitudes in each stripe domains are present, in parallel with increasing in-plane components of the magnetization. At larger fluences, XMCD-PEEM images reveal the new partially ordered stripe domain structures characterized by a new local magnetic domain pattern showing organized pattern at the micrometer scale. We provide a local real space description of these new magnetic domains and show that they are independent from the initial stripe configurations. We confirm the mechanism inducing such reorganisation of stripe domains in CoPd films, considering the recovery of the magnetic anisotropy during the heat dissipation under an external Oersted field.

**Figures and Tables of the manuscript by Lopes-Flores et al.**

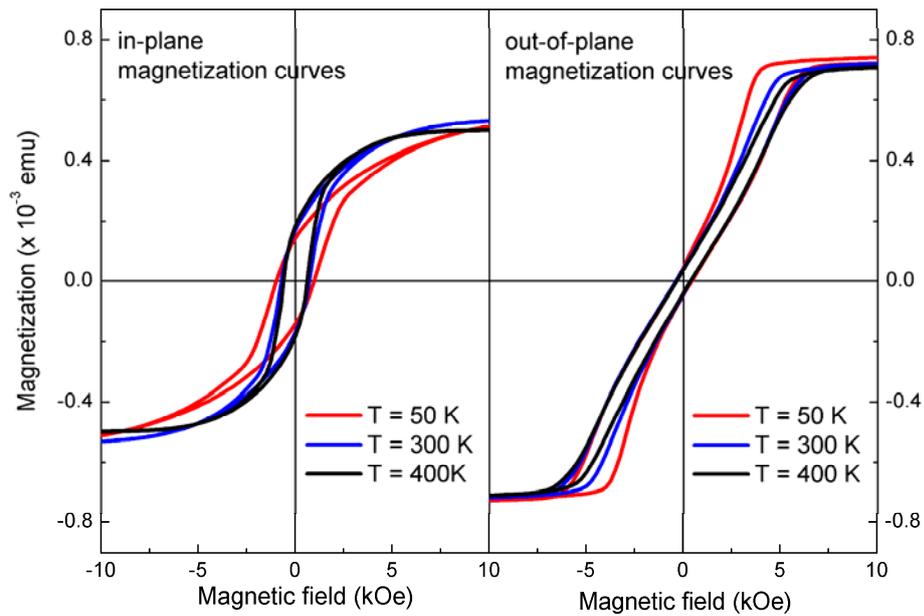

**Figure 1:** SQUID magnetometry measurements showing the magnetization curves of the sample at different temperatures along the in-plane (left), and out-of-plane direction (right).

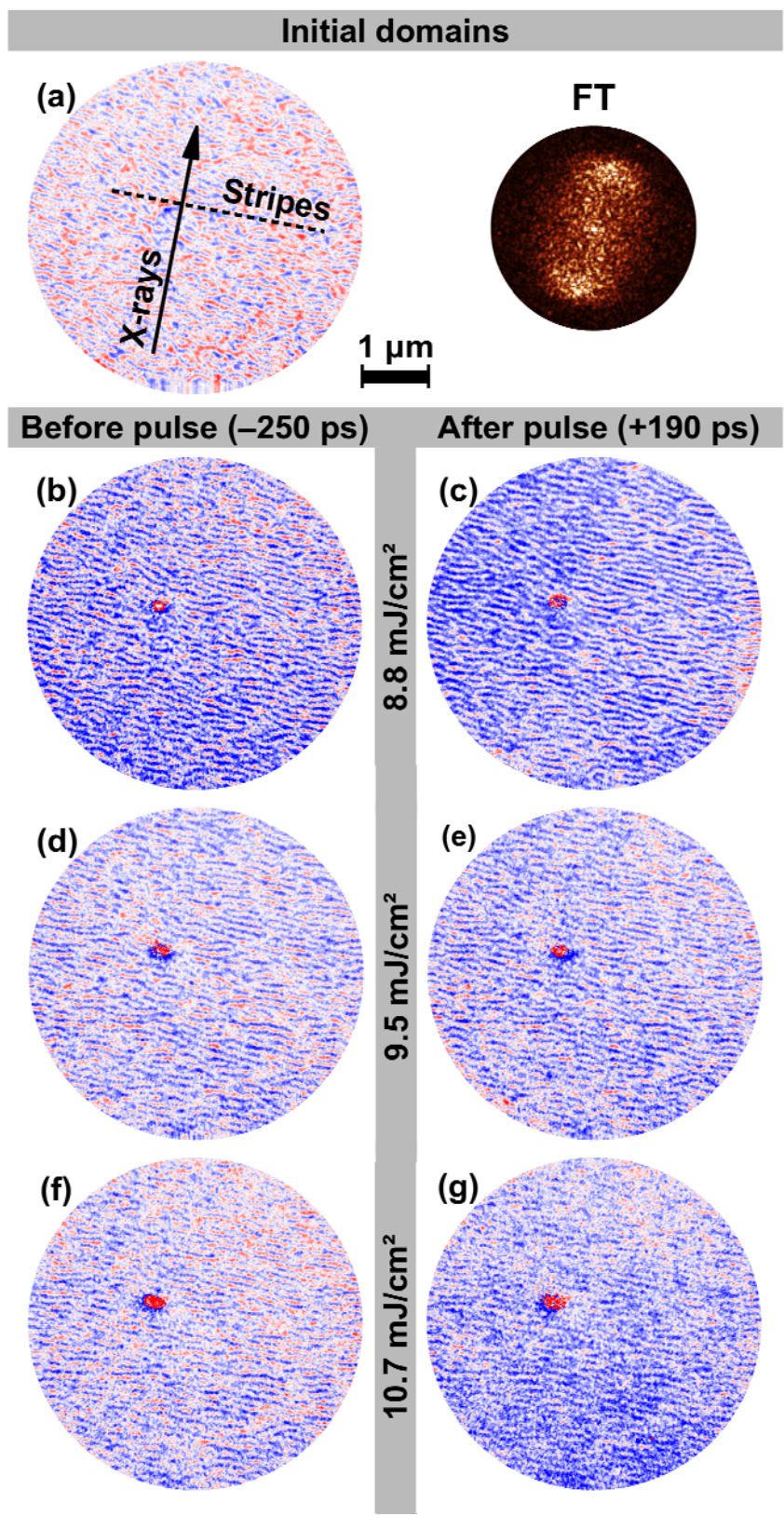

**Figure 2:** Time-resolved XMCD-PEEM images of the sample with stripes oriented 90º from the X-

ray incidence, without any laser applied at room temperature (a), with an IR fluence of 8.8 mJ/cm$^2$ at t = -250 ps (b), with an IR fluence of 8.8 mJ/cm$^2$ at t = +190 ps (c), with fluence of 9.5 mJ/cm$^2$ at t = -250 ps(d), with with fluence of 9.5 mJ/cm$^2$ at t = +190 ps (e), with fluence of 10.7 mJ/cm$^2$ at t = -250 ps (f) and with fluence of 10.7 mJ/cm$^2$ at t = +190 ps (g). The FOV of the XMCD-PEEM images is 5 µm. In (a) the initial XMCD-PEEM image (left) is compared with the corresponding Fourier Transform (FT) image (right).

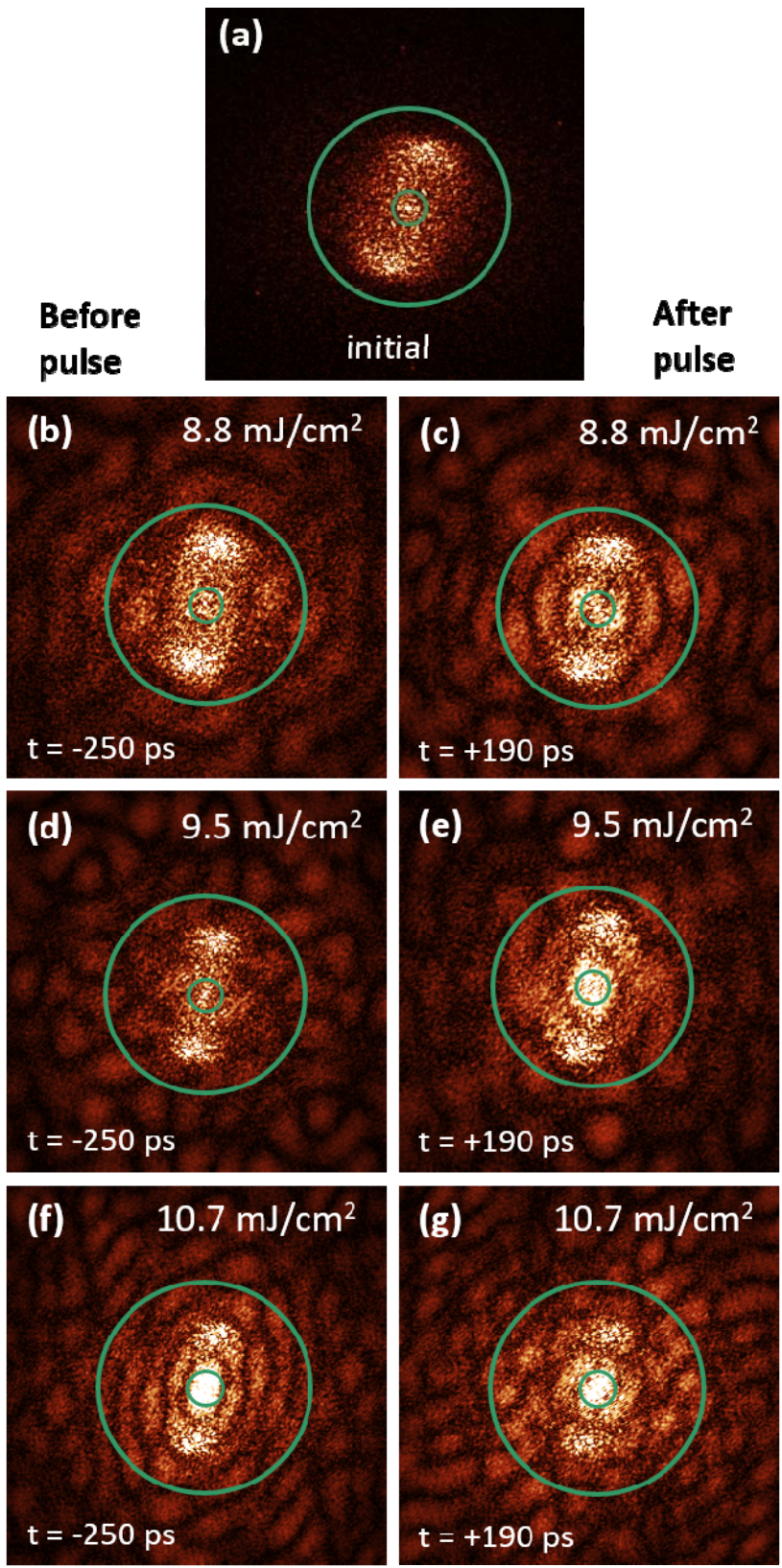

**Figure 3:** Time-resolved Fourier Transform (FT) of the XMCD-PEEM images as given in figure 2. The green circles show the limits used for the azimuth integration defining the intensity distributions I(Φ).

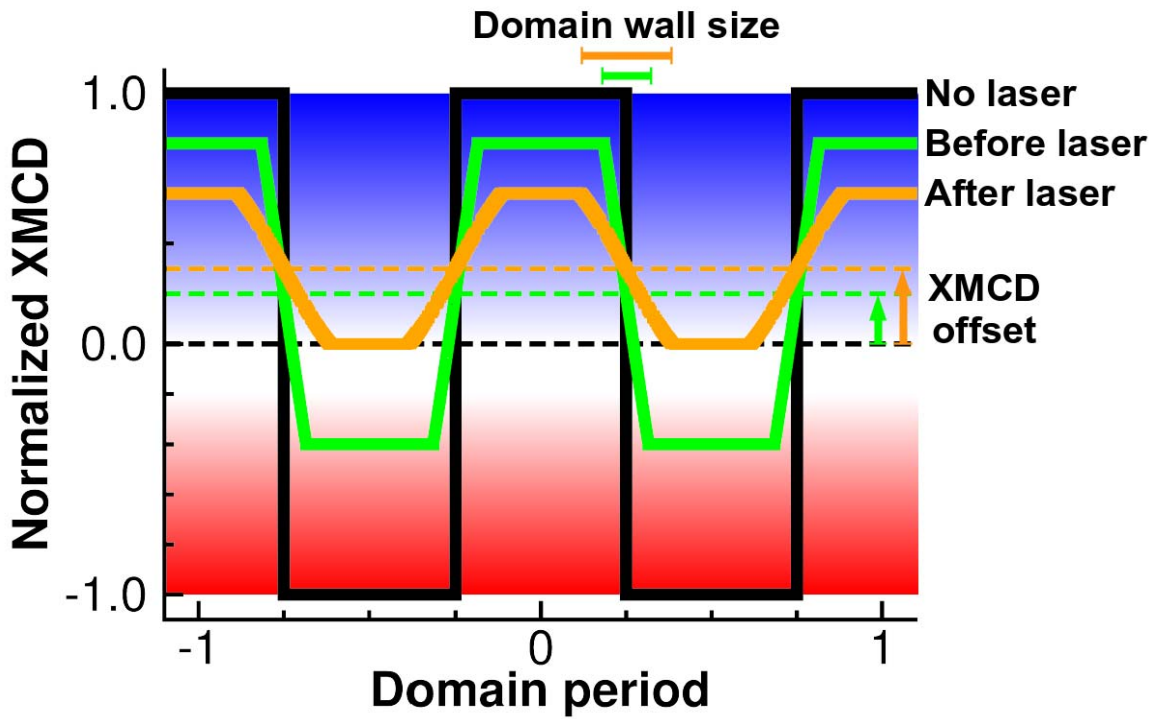

**Figure 4:**
Qualitative representation of the profile line intensities for 3 different XMCD-PEEM configurations (i.e. figure 2 a, b, and c). It shows 3 line scans perpendicular to the stripe direction, as observed in our XMCD-PEEM images for 3 different stripe domain configurations: Without laser (black line), with laser at a delay of -250 ps (green line) and with laser at a delay of +190 ps (orange line). Continuous lines (black, green, and orange) evidence the local magnetization across the domain walls. An XMCD offset shift towards higher XMCD values is shown by green and orange dashed lines for both delays -250 ps and +190ps. The blue-white-red color scale is related to the color code at the X-PEEM images.

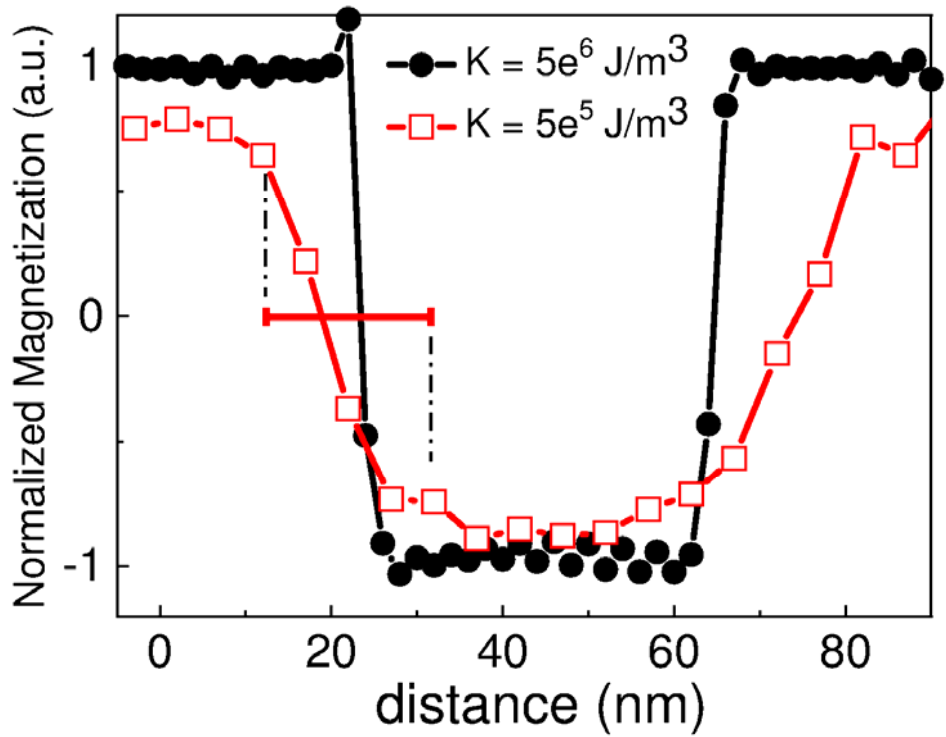

**Figure 5**: Two simulated magnetic line profiles by OOMMF micromagnetic simulations. It shows two different magnetic line profiles calculated by using realistic values of the magnetic anisotropy constants defining the system at room temperature and at an elevated temperature, $K_m$=5e6 J/m$^3$ and $K_m$=5e5 J/m$^3$, respectively. The OOMMF simulated line profiles evidence the change of the domain wall between two up and down domains. The simulations show that at lower anisotropy (higher temperature) the domain wall thickness is broader than at high anisotropy (15 nm instead of ~3 nm).

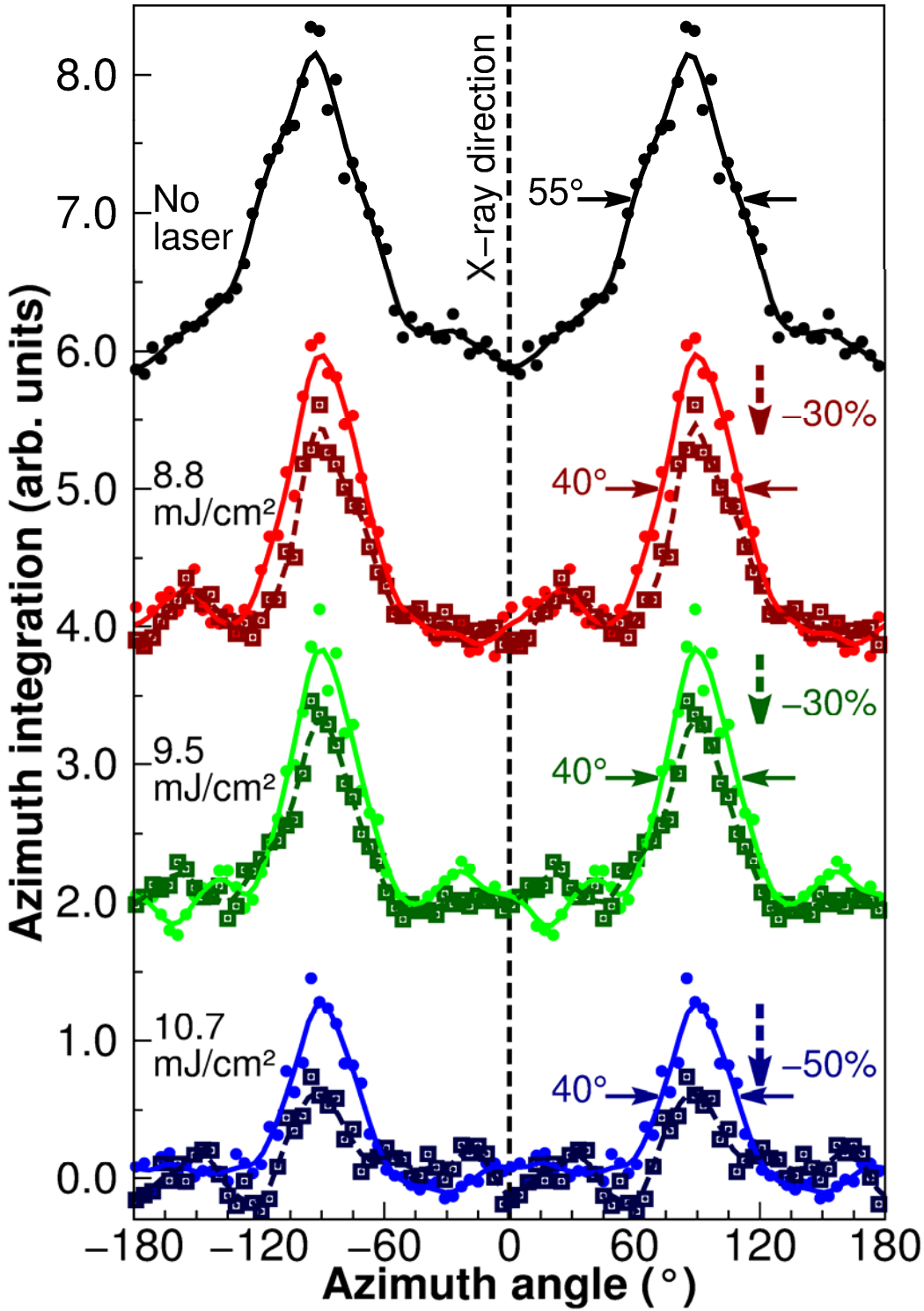

**Figure 6:** Azimuthal intensity distributions I(Φ) extracted from the Fourier transform diagrams in Figure 3, delimited by the green circles. The graphs are grouped in pairs, obtained with the same laser power (0, 8.8, 9.5, 10.7 mJ/cm$^2$). The lines are guides to the eyes. Dots and their corresponding full lines come from the images taken at t = -250 ps. Squares and dashed lines come from the images taken at t = +190 ps. The vertical dashed arrows show the intensity reduction, accounting for the reduction of the XMCD contrast in the XMCD-PEEM images. The horizontal full arrows show the width at half maximum of the peaks.

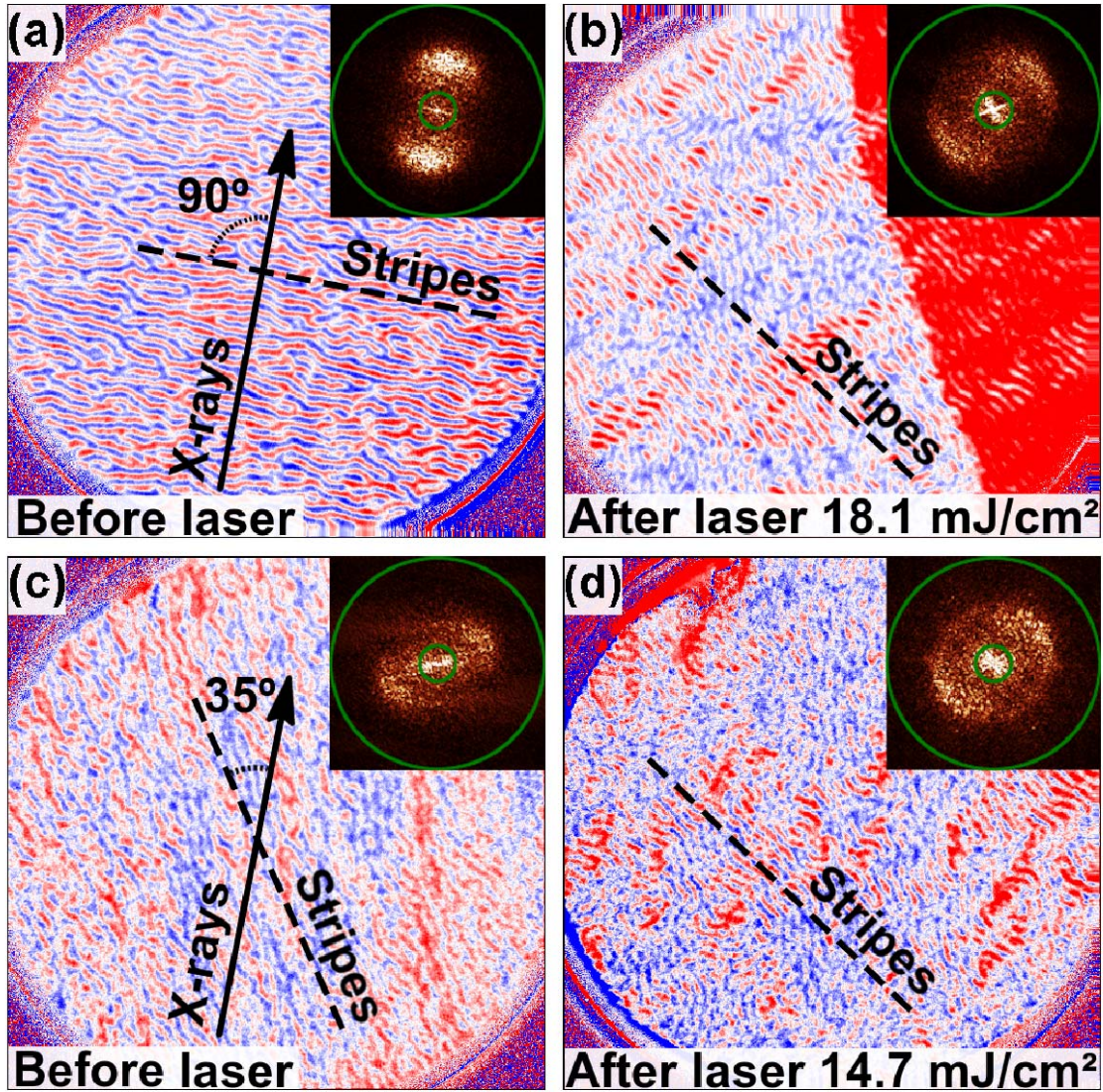

**Figure 7:** XMCD-PEEM images of the sample (a) with stripes oriented at 90º from the X-ray incidence, at room temperature (no laser applied) (b) at the same position and orientation as (a) but after pump excitation at 18.1 mJ/cm$^2$; (c) with stripes oriented at 35º from the X-ray incidence, at room temperature (no laser applied) (d) at the same position and orientation as (c) but after pump excitation at 14.7 mJ/ cm$^2$. The FOV of the XMCD-PEEM images is 5 μm. The insets in the XMCD-PEEM images show the corresponding FT of the images, which account for the preferential orientation of the stripes. The green circles show the limits used for the azimuth integration defining the intensity distributions I(Φ).

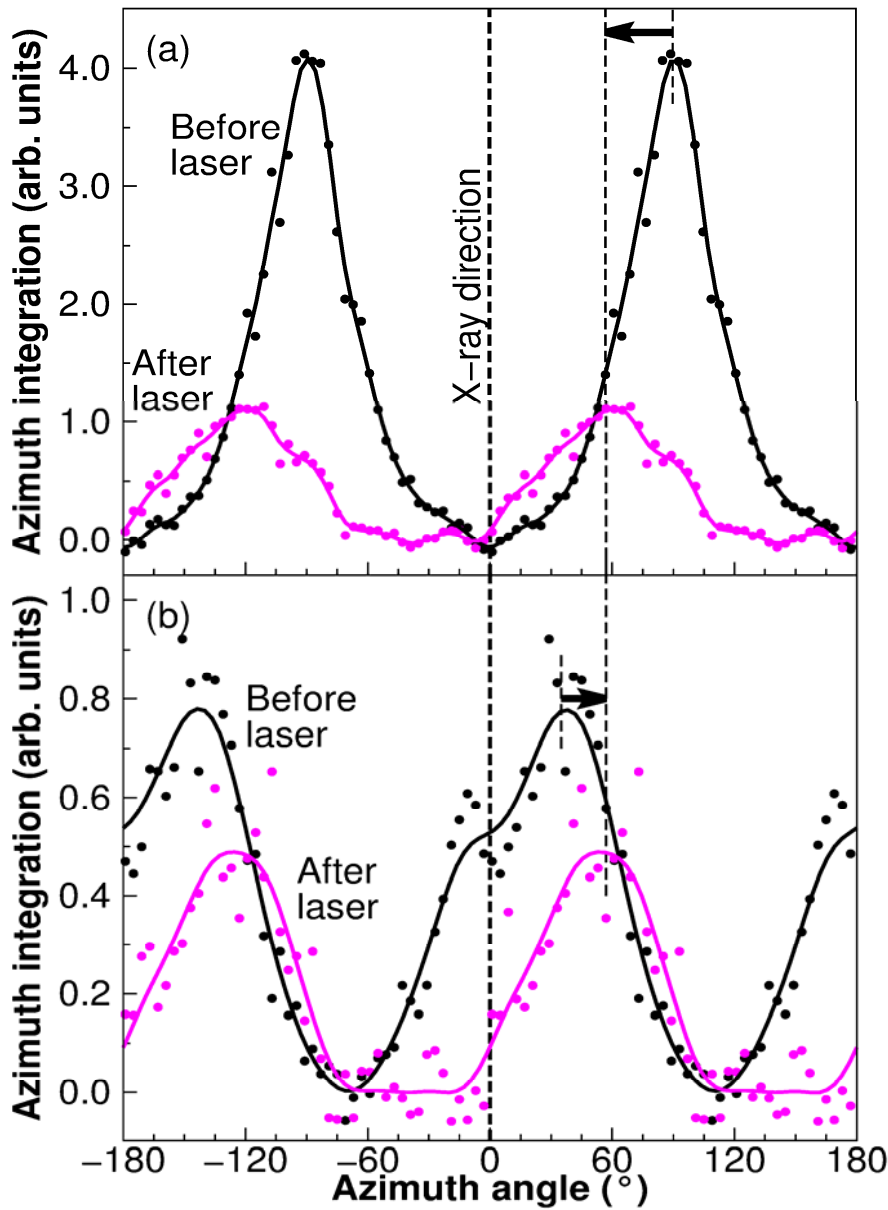

**Figure 8:** Azimuthal integration of the Fourier transform diagrams for the corresponding XMCD-PEEM images, (a) for initial orientation of the stripes of 90º showing in red the effect of the laser pump excitation at 18 mJ/cm², and (b) for initial orientation of 35º 90º showing in blue the effect of the laser pump excitation at 15 mJ/cm². The arrows show the reorientation direction to the new stripes at ~ 60º.

|  | XMCD @ t = -250 ps | XMCD @ t = +190 ps |
| --- | --- | --- |
| No Laser | 0.8 % |  |
| 8.8 mJ/cm$^2$ | 9.7 % | 11.3 % |
| 9.5 mJ/cm$^2$ | 5.2 % | 6.7 % |
| 10.7 mJ/cm$^2$ | 3.7 % | 5% |

**Table 1:** Mean XMCD values of the XMCD-PEEM images extracted from Figure 2

*Acknowledgments*:

We are indebted for the scientific discussions with N. Bergeard from IPCMS Strasbourg, concerning the OOMMF micromagnetic simulations. The authors are grateful for financial support received from the following agencies: the French CNRS-PICS program, the EU Contract Integrated Infrastructure Initiative I3 in FP6 Project No.R II 3CT-2004-50600008 and the European Community's Seventh Framework Programme (FP7/2007-2013) under grant agreement n.°312284."
.